\documentstyle[12pt]{article}
\newcommand{\be}{\begin{equation}}
\newcommand{\ee}{\end{equation}}
\newcommand{\bea}{\begin{eqnarray}}
\newcommand{\eea}{\end{eqnarray}}

\begin{document}
 \begin{titlepage}

\begin{flushright}
CERN-TH.7155/94
\end{flushright}
\vspace{20 mm}

\begin{center}
{\huge Two-Dimensional String Theory,}

\vspace{5mm}

{\huge  Topological Field Theories }

\vspace{5mm}

{\huge and the Deformed Matrix Model}

\end{center}

\vspace{10 mm}

\begin{center}
Ulf H. Danielsson\\
Theory Division, CERN, CH-1211 Geneva 23, Switzerland

\end{center}

\vspace{2cm}

\begin{center}
{\large Abstract}
\end{center}
In this paper the $c=1$ string theory is studied from the point
of view of topological field theories. Calculations are
done for arbitrary genus. A change in the prescription is
proposed, which reproduces the results of the $1/x^2$
deformed matrix model. It is proposed that the deformed matrix
model is related to a D-series Landau-Ginzburg
superpotential.

\vspace{2cm}
\begin{flushleft}
CERN-TH.7155/94 \\
January 1994
\end{flushleft}
\end{titlepage}
\newpage

\section{INTRODUCTION}

The tremendous success of the matrix models in describing two
dimensional quantum gravity and low-dimensional string theory is
in sharp contrast with the difficulties of the continuum approach.
Recent developments connecting the $c=1$ model to an
unorthodox
Landau-Ginzburg model are beginning to change this. This
promises to give new insights into the structure, and physics, of
two-dimensional string theory.

Such an improved understanding is important for several reasons.
The target space physics of the $c=1$ matrix model is very badly
understood. It has been shown, see [1-6], that a deformation of
the matrix model potential gives a new model, presumably also a
model of a string moving in a two-dimensional target space. It has
been conjectured that this model is the black hole of ref. \cite{wit}.

In this context it is important to come with some clarifications.
In the work of \cite{muva} it has been suggested that the ordinary
$c=1$ model can be thought of as describing a black hole.
This is a consequence of the Landau-Ginzburg description
and its association with the coset model
\be
\frac{SU(2)_{k=-3} \otimes [b,c]}{U(1)} ,
\ee
which was derived in \cite{muva}.
However, as pointed
out in \cite{muva}, this is not the
black hole described by
Witten, which instead is given by the coset
\be
\frac{SU(2)_{k=-9/4} }{U(1)}\otimes [b,c]  .
\ee

In this paper I will suggest a modification of the super-potential
prescription for $c=1$, which reproduces the results of the
deformed matrix model. The hope is that this will give further
insight into the physics of the deformed matrix model and the
target space physics of the two-dimensional string.

In Section 2 I will quickly go through the very basics of the
connection between Landau-Ginzburg models and topological
field theories. In Section 3 the techniques are generalized to $c=1$.
Section 4 is devoted to higher genus, and Section 5 contains
a comparison with more traditional $c=1$ matrix model calculations.
Then, in Section 6, I propose a generalization of the
construction, which is capable of describing the deformed matrix model.
This is found to work also at higher genus. I also show that this
implies that the deformed matrix model is associated with a D-series
superpotential.

\section{LANDAU-GINZBURG POTENTIALS AND GELFAND-DIKII EQUATIONS}

I will begin by recalling some of the basic building blocks in the
description of topological string theory. In particular the relations
between the KdV hierarchy and the Landau-Ginzburg potential.

By starting with the two-dimensional minimal $N=2$ superconformal field
theories it is possible to obtain a sequence of
topological field theories, see e.g. \cite{dvv}. These
$N=2$ models can be
described by super potentials $W(X)$. The super  potentials are
classified according to the A-D-E scheme. For instance, one has the
A-series with the potentials $W(X)=X^{k+2}$. They correspond to
models with central charge $d=\frac{k}{k+2}$ with $k$ positive
integer. These models can also be realized as
$\frac{SU(2)_{k}}{U(1)}$ coset models, see \cite{coset}. As shown in
\cite{dvv} there is also a direct relation between the super potentials
and the KdV equations of the $c<1$ matrix models,
which I will quickly go through below.
A nice reference is \cite{wit3},
from where I will borrow some of the notation.

The crucial observation in \cite{dvv} was that the superpotential,
and its perturbations, can be directly identified with the
KdV differential operator, i.e.
\be
W=D^{k+2}
 -\sum _{i=0}^{k} u_{i}(x)D^i  ,
\ee
in the case of the A-series.
I have replaced $X$
with $D=\frac{i\hbar}{\sqrt{k+2}} \frac{\partial}{\partial x}$ and
$x$ is
the coupling of the puncture operator.
The
$u_{i}(x)$ are determined through the flow equations
\be
i\frac{\partial W}{\partial t_{m,n}} =
\frac{(k+2)^{m+1/2} (-1)^{m}}
{(n+1)(n+k+3)...(n+1+m(k+2))} \left[
W_{+}^{m+\frac{n+1}{k+2}} , W\right]
\ee
with $t_{0}=x$;
$t_{m,n}$ is the coupling of $\sigma _{m} (\phi _{n})$, i.e. the
$m$th descendant of the primary $\phi _{n}$. The $+$-index means
keeping only positive powers of $D$.
Restricting to the
flow due to primaries one gets
\be
i\frac{\partial W}{\partial t_{n}} =\frac{(k+2)^{1/2}}{n+1}\left[
W_{+}^{\frac{n+1}{k+2}} , W\right]  .
\ee
I have put $t_{n}=t_{0,n}$. The correlation functions of
topological field theory are then defined through
\be
\langle \phi _{j} \rangle = \frac{\partial F}{\partial t_{j}} =
\frac{(k+2)^2}{(j+1)(j+k+3)} res(W^{1+\frac{j+1}{k+2}})    ,
\ee
where 'res' picks out the coefficient of $D^{-1}$. Higher-order
correlation functions are obtained by taking more derivatives and
using the flow equations. One point functions of descendants are
given by
\be
\langle \sigma _{m} (\phi _{j} ) \rangle =
\frac{(-1)^{m}(k+2)
^{m+2}}{(j+1)(j+k+3)...((m+1)(k+2)+j+1)}
  res(W^{1+m+\frac{j+1}{k+2}})       .
\ee

Let me give an example of a correlation function.
In the classical ($\hbar \rightarrow 0$) limit where
commutators are replaced by Poisson brackets, the three-point
function for three primaries is given by
$$
\frac{\partial ^3 F}{\partial t_{m_{1}} \partial t_{m_{2}}
\partial t_{j}}
$$
$$
=
\frac{(k+2)^2}{(j+1)(j+k+3)(m_{1}+1)(m_{2}+1)} res\left(
\left\{
W_{+}^{\frac{m_{1}+1}{k+2}},\left\{ W_{+}^{\frac{m_{2}+1}{k+2}},
W^{1+\frac{j+1}{k+2}}\right\}  \right\}     \right)
$$
\be
=\frac{k+2}{(j+1)(m_{1}+1)(m_{2}+1)} res\left(
\frac{\partial _{D}
W_{+}^{\frac{m_{1}+1}{k+2}}\partial _{D} W_{+}^{\frac{m_{2}+1}{k+2}}
\partial _{D} W^{\frac{j+1}{k+2}}}{\partial _{D} W} \right) .
\ee
I have used $\frac{\partial}{\partial x} W_{+}^{\frac{m+1}{r}} =0$,
for $m \leq k$.
At zero coupling  $W_{+}^{\frac{m+1}{k+2}}=D^{m+1}$, this
implies the standard result
\be
\frac{\partial ^3 F}{\partial t_{m_{1}} \partial t_{m_{2}}
\partial t_{j}} = \delta _{m_{1}+m_{2}+j,k}  .
\ee
Higher-point correlation functions are obtained analogously.
This illustrates the connection between the KdV-operator
the Landau-Ginzburg models.
In the following I will often be careless and generally refer to
$W$ as the superpotential, even when the argument is a
differential operator $\frac{\partial}{\partial x}$.

\section{THE $c=1$ MODEL FROM A $D^{-1}$ POTENTIAL}

In \cite{wit2} it was found that if $k=-3$ then the topological theory
computes the Euler characteristics of the moduli space of
Riemann surfaces. Since this is also accomplished by the
$c=1$ matrix model at the self dual radius $R=1$ (in units
where $\alpha ' =1$), \cite{diva},
it is reasonable to expect that this model
{\it is} the $c=1$ string at $R=1$. In \cite{muva} this conjecture
was proven by an explicit construction of a twisted $N=2$
theory starting with $c=1$. The supercurrents were obtained
by using the $b-c$ ghost system. Hence one is lead to consider
models with superpotential $W=X^{-1}$.

In this and the next few sections I will illustrate the $k=-3$ case
by some explicit examples and calculations, including higher genus.
This is intended as a warm-up for Section 6 where,
finally, I propose a way of extending the formalism to the case of
the deformed matrix model.

Now, to understand how all of this works, we must adapt the
formalism of the previous section to the case $D^{-1}$. This is
not completely straightforward and several papers have been
devoted to this subject \cite{muva,han,mugh}.

I begin by specifying the superpotential $W$. I restrict myself to
the small phase space, i.e. where all couplings to descendants are
put to zero. The superpotential is then
\be
W =\mu D^{-1} - x +\sum _{i=2}^{\infty} t_{i} D^{i-1} ,
\label{sup}
\ee
where $D=-i\hbar \frac{\partial}{\partial x}$. I will sometimes
keep the $\hbar$ for clarity but at the end it will always
be put to 1,
or, more precisely, absorbed into $\mu$.
It is important that $W$ is only linear in the couplings $t_{i}$.
This implies that the primaries
\be
\phi _{m} = \frac{\partial W}{\partial t_{m}} = D^{m-1}
\label{prim}
\ee
do not flow.

The primaries $\phi _{m}$ are identified with tachyons, $T_{m}$, with
positive momenta $m>0$. The negative-momentum tachyons,
$T_{-m}$, correspond to descendants $\sigma _{m} (\phi _{0})$ of
the cosmological constant operator, i.e. $D^{-1}$. One might note that the
puncture operator, in the sense of topological field theory, is the
first special tachyon, $T_{1}$. This was first observed in \cite{kita}.

The one-point function of a negative-momentum tachyon is
\be
\langle T_{-p} \rangle = \frac{1}{p(p+1)} res(W^{p+1})   .
\label{1t}
\ee
{}From this, the expression for $W$ and the identification of the
positive-momentum tachyons as the primaries (\ref{prim}), one can
easily verify the matrix model results for the tachyon correlation
functions. Indeed,
$$
\frac{1}{p(p+1)} res \left(
\mu D^{-1} -x +\sum _{i=2}^{\infty} t_{i}
D^{i-1} \right)  ^{p+1}
$$
\be
= \frac{1}{p(p+1)} \sum_{\sum _{i=1}^{l} m_{i}n_{i} =p}
t_{n_{1}}^{m_{1}} ... t_{n_{l}}^{m_{l}}
\mu ^{p+1-\sum _{i=1}^{l}m_{i}} \frac{(p+1)p...(p+2-
\sum _{i=1}^{l}m_{i})}{m_{1}!...m_{l}!}    .
\ee
Hence
\be
\langle T_{p_{1}}...T_{p_{N-1}}T_{-p} \rangle =
\frac{\partial}{\partial t_{p_{1}}}...
\frac{\partial}{\partial t_{p_{N-1}}} \langle T_{-p} \rangle =
(p-1) ... (p-N+3) \mu ^{p-N+2}          .
\ee
In \cite{han,mugh}, correlation functions of more than one descendant,
i.e. negative-momentum tachyon, are considered. This implies
the presence of contact terms that need special care. I will not
consider these complications. I will instead restrict myself to
the simpler case where no contact terms are needed.

The main subject of the next section is to show that these
calculations can be extended to higher genus. The claim is that if
one remembers that
$D=-i\hbar \frac{\partial}{\partial x}$, where $x$ is
the coupling of the $T_{1}$ tachyon and if one
defines the residue as the
coefficient of $D^{-1}$ after that all the $x$'s have been
commuted to the left, then the higher genus results of the
matrix model are reproduced.

Rather than proving this in the same way as above for genus zero,
I will do it in a way that clearly illustrates the close connection
with the matrix model. In fact, it will turn out that the
calculations are formally identical to calculations
of tachyon correlation functions done in [17-19].
I will use the property that
\be
\frac{\partial}{\partial t_{p}} = \frac{1}{i\hbar}
\left[ T_{p} , \cdot \right]      ,
\label{derkom}
\ee
to replace the $t_{p}$ derivatives with commutators. This will be the
subject of the next section.

\section{CORRELATION FUNCTIONS AT HIGHER GENUS}

\subsection{The Algebraic Structure}

Let me reconsider the calculation, in the previous section, of the
tachyon correlation function, but now using relation
(\ref{derkom}).
This implies
\be
\langle T_{k_{1}}... T_{k_{N-1}}T_{-k_{N}}\rangle
= \frac{(i\hbar)^{1-N}}{k_{1}...k_{N}}
res(\left[  T_{k_{1}}, \left[ T_{k_{2}},\left[ ... \left[ T_{k_{N-1}} ,
T_{-k_{N}}
\right] ...\right] \right. \right. )
\ee
where
\be T_{k} = D^{k}
\ee
and
\be
T_{-k} =\frac{1}{k+1} (\mu D^{-1} -x)^{k+1}
\ee
for $k>0$. What is the algebra generated by the tachyons?
On the sphere one finds, for the Poisson brackets,
\be
\{ T_{k} , T_{-l} \} = kl T_{k-1}T_{-l+1}
\ee
and in general
$$
\frac{1}{i\hbar} \left[ T_{k}, T_{-l}\right] =\left\{ T_{k}, T_{-l}
\right\}
_{M} =  \frac{2\mu}{\hbar} \sin \frac{\hbar}{2\mu}
\left(
\frac{\partial}{\partial
D_{2}} \frac{\partial }{\partial x_{1}} - \frac{\partial}{\partial
D_{1}} \frac{\partial }{\partial x_{2}} \right)
T_{k} T_{-l}
$$
\be
= kl(T_{k-1}T_{-l+1})_{W} +
\frac{\hbar ^{2}}{24\mu ^{2}} k(k-1)(k-2)l(l-1)(l-
2)(T_{k-3}T_{-l+3})_{W} +...
\label{walg}
\ee
where $\{ ,\} _{M}$ denotes the Moyal bracket \cite{moyal}, and the
subscript $W$ means Weyl ordering, (i.e. a sum over all possible
orderings with equal weight). This is precisely the $W_{\infty}$
algebra of the $c=1$ matrix model. In the matrix model, see
\cite{wit4}, the generators are
\be
W_{J,m} = (\lambda +p)^{J+m} (\lambda -p )^{J-m} ,
\ee
obeying
\be
\left[ W_{J_{1},m_{1}}, W_{J_{2},m_{2}}\right] =4(m_{2}J_{1}-
m_{1}J_{2}) W_{J_{1}+J_{2}-1, m_{1}+m_{2}} +...
\ee
with $\hbar$ corrections given by the Moyal bracket if the
operators are defined using Weyl-ordering, \cite{min2,avhand,spenta}.
So, we have seen that the algebraic structure of the model
defined in the previous section is identical to that of the
matrix model. This is true to for all genera.
In the next section I will
verify that the correspondence is also true for the one-point
functions. The knowledge of the algebra and its higher genus
corrections, together with the knowledge of the one-point
functions, make it possible to calculate the correlation
functions at arbritrary genus.

\subsection{The One-Point Functions}

In this section I will calculate the one-point functions
\be
res\left( (D^{r}(\mu D^{-1}-x)^{r+1} )_{W} \right)   .
\ee
These correspond to one-point functions of $W_{r,0}$. The residue is
defined as the coefficient of the $D^{-1}$ term
when all the $D$'s have been commuted to the right.

I will need the formulae
\be
(q^{m}p^{n})_{W} = 2^{-m} \sum _{l=0}^{m} \left( \begin{array}{c}
m \\ l
\end{array} \right) q^{m-l} p^{n} q^{l}
\ee
and
\be
\left[ p^{n},q^{m} \right] = \sum _{k=1}^{min(n,m)}
\left( \begin{array}{c}
m \\ k
\end{array} \right)
\left( \begin{array}{c}
n \\ k
\end{array} \right)  k!
(-i\hbar)^{k}q^{m-k} p^{n-k}   ,
\ee
given $[p,q] =-i\hbar$. A derivation of the first formula, given the
definition of Weyl-ordering, can be found for instance
in \cite{lee}. It
is only valid for positive $n$ and $m$. In the second formula, however,
one can allow either $m$ or $n$ to be negative.
These combine to
\be
(q^{m}p^{n})_{W}=
\sum _{k=0}^{min(n,m)}
2^{-k}
\left( \begin{array}{c}
m \\ k
\end{array} \right)
\left( \begin{array}{c}
n \\ k
\end{array} \right)  k!
(-i\hbar)^{k}q^{m-k} p^{n-k}   ,
\ee
valid only for $n$ and $m$ positive.
Here $\sum _{l=0}^{n} \left( \begin{array}{c}
n\\ l
\end{array} \right) =2^{n}$ has been used. Since $D$ and
$-\mu D^{-1} +x$ obey the same algebra as $p$ and $q$, it follows
that
$$
\left( D^{r}(\mu D^{-1}-x)^{r+1}\right) _{W}
$$
\be
=
\mu ^{r+1} \sum _{k=0}^{r} 2^{-k}
\left( \begin{array}{c}
r+1 \\ k
\end{array} \right) \frac{r!}{(r-k)!} (\frac{i}{\mu}) ^{k}
(\mu D^{-1}-x)^{r+1-k}
D^{r-k}         .
\ee
Let me now commute $x$ to the left in the powers of $\mu D^{-1} -x$
and then put $x=0$. This gives
$$
\mu D^{-1}-x \rightarrow \mu D^{-1}     ,
$$
$$
(\mu D^{-1}-x )^{2}\rightarrow \mu D^{-1} (\mu D^{-1}-x) \rightarrow
\mu ^{2} \left(
1+\frac{1}{i\mu}
\right)
D^{-2}
$$
and in general
\be
(\mu D^{-1}-x )^{k} \rightarrow \left( 1
+\frac{1}{i\mu} \right)
...\left( 1+(k-1)
\frac{1}{i\mu} \right) D^{-k}    .
\ee
So the result is
$$
res\left( D^{r} (\mu D^{-1} -x)^{r+1} \right) _{W}
$$
$$
= \mu ^{r+1}\sum
_{k=0}^{r} 2^{-k} \left( \begin{array}{c}
r+1 \\ k
\end{array} \right) \frac{r!}{(r-k)!} \left(
1+\frac{1}{i\mu} \right)
...\left( 1+
\frac{r-k}{i\mu} \right)
\left( \frac{i}{\mu} \right) ^{k}
$$
\be
= \sum _{k=0}^{r} 2^{-k} \left( \begin{array}{c}
r+1 \\ k
\end{array} \right) \frac{r!}{(r-k)!} i^{r-1} (-1)^{r-k}
(i\mu )_{r-k+1}.
\label{hr}
\ee
where $(x)_{n} = x(x+1)...(x+n-1)$.
An expansion in genus is easily obtained using Stirling numbers of
the first kind, $S_{n}^{(m)}$, as defined in \cite{abr}:
\be
x(x-1)...(x-n+1)=\sum _{m=0}^{n} S_{n}^{(m)} x^{m}  .
\ee
We then get
\be
\sum _{k=0}^{r} 2^{-k} \left( \begin{array}{c}
r+1 \\ k
\end{array} \right) \frac{r!}{(r-k)!} i^{r+1} \sum _{m=0}^{r-k+1}
S_{r-k+1}^{(m)} (-i\mu )^{m}       .
\ee
The first few terms in the series are
\be
\mu ^{r+1} \left(
1-\frac{(r+1)r(r-1)}{12\mu ^{2}} +\frac{(r+1)r(r-1)(r-2)(r-3)}
{5760\mu ^{4}}(20r-8) +... \right) .
\ee

\subsection{An Example}

To illustrate the above formulae, consider the two-point function
$\langle T_{k}T_{-k}\rangle$ to genus one. First use the $W_{\infty}$
algebra of (\ref{walg}) to obtain
$$
\langle T_{k}T_{-k} \rangle = \frac{1}{i\hbar k^{2}} res(\left[T_{k} ,
T_{-k} \right] )
 = res\left( D^{k-1} \frac{(\mu D^{-1} -x)^{k} }{k}\right) _{W}
$$
\be
+
\frac{1}{24\mu ^{2}} (k-1)^{2}(k-2)^{2}
res\left( D^{k-3} \frac{(\mu D^{-1} -x)^{k-2}}{k-2} \right) _{W} +...
\ee
Then use the residue formula of the previous section to obtain the
answer
\be
\mu ^{k} \left(\frac{1}{k} -\frac{(k-1)(k-2)}{12\mu ^{2}} +
\frac{(k-1)^{2}(k-2)}{24\mu ^{2}} \right)
=\frac{\mu ^{k}}{k} - (k-1)(k^{2}-k-2)\frac{\mu ^{k-2}}{24}
\ee
for the two-point to genus 1. This agrees with the result of \cite{klow}.

\section{A COMPARISON WITH THE MATRIX MODEL}

I will now compare the above calculational rules and results
with what is obtained in the matrix model. I will use the
techniques developed in \cite{min2,avhand}. There it was shown how
perturbation theory leads to the formula
\be
\langle PT_{k_{1}}...T_{k_{N}} \rangle \sim
\langle P\left[ T_{k_{1}},\left[ T_{k_{2}},
...\left[
 T_{k_{N-1}},T_{k_{N}}\right] ... \right] \right] \rangle  \label{corr}
\ee
up to the factorized external leg factors;  $k_{1}$ through $k_{N-1}$
are positive  while $k_{N}$ is negative.
The right-hand-side two-point function is given by
\be
\langle PW\rangle =-\frac{1}{\pi} {\cal I} \sum_{n=0}^{\infty}
\frac{\langle n \| W \| n \rangle}{E_{n}-\mu};\label{pw}
\ee
$n$ labels the one-particle states in the matrix model
harmonic potential and $E_{n}= \frac{2n+1}{2i}$ at $\alpha ' =1$.
I have already, in
eq. (\ref{walg}), written down the algebra and its higher genus
($\hbar$) corrections. To complete the matrix model calculation
it is necessary to calculate
\be
\langle (H^{r})_{W} P \rangle     .
\ee
The easy object to calculate is
\be
\langle H^{r} P \rangle = \mu ^{r} \log \mu          .
\label{hp}
\ee
This follows immediately from the formal equality
\be
\sum_{n=0}^{\infty} \frac{f(n)}{n+z}  = f(-z) \sum _{n=0}^{\infty}
\frac{1}{n+z}                                    ,
\ee
if one only keeps the terms with a $\log \mu$. This is legitimate
since, at a fixed positive power of $\mu$, these terms will dominate.
In (\ref{hp})
$H=\frac{1}{2}(aa^{\dagger} + a^{\dagger}a)$ is Weyl-ordered.
Henceforth I will drop all $\log \mu$'s and external pole factors.
My normalizations will be those of collective field theory.

Following \cite{avhand} I will now calculate $(H^{r})_{W}$. Consider
\be
(a^{r}a^{\dagger r})_{W} = 2^{-r} \sum _{l=0}^{r} \left(
\begin{array}{c}
r \\ l
\end{array} \right) a^{\dagger r-l} a^{r} a^{\dagger l}   .
\ee
Take the expectation value of this in the harmonic oscillator
state $n$:
\be
2^{-r} \sum _{l=0}^{r} \left( \begin{array}{c}
r \\ l
\end{array} \right) (n+l)...(n+l-r+1)       .
\ee
Use $\langle n | H | n \rangle = n +\frac{1}{2}$ to deduce
\be
(a^{r}a^{\dagger r})_{W} = 2^{-r} \sum _{l=0}^{r} \left(
\begin{array}{c}
r \\ l
\end{array} \right) \left( H-\frac{1}{2}+l\right)
...\left( H-\frac{1}{2}+l-r+1\right)
\ee
and hence
\be
\langle (H^{r})_{W} P\rangle = (2i)^{-r} \sum _{l=0}^{r} \left(
\begin{array}{c}
r \\ l
\end{array} \right) \left(
i\mu -\frac{1}{2}+l\right)
...\left( i\mu -\frac{1}{2}+l-
r+1 \right)
\label{hwp}  .
\ee
This result is valid for the uncompactified case, i.e. $R=\infty$,
to obtain the result for finite radius $R$, one should act with
\cite{klow}:
\be
\frac{\frac{i}{R}\frac{\partial}{\partial \mu}}
{e^{\frac{i}{2R}\frac{\partial}{\partial \mu}} - e^{ -
\frac{i}{2R}\frac{\partial}{\partial \mu}}}   .
\ee
That is, at the self-dual radius $R=1$:
\be
\langle (H^{r})_{W} \rangle _{R=1} =
\frac{i}{e^{\frac{i}{2}\frac{\partial}{\partial \mu}} - e^{ -
\frac{i}{2}\frac{\partial}{\partial \mu}}}
\langle (H^{r})_{W} P\rangle     .
\ee
I will now prove that, indeed,
\be
i\langle (H^{r})_{W} P\rangle =
(e^{\frac{i}{2}\frac{\partial}{\partial \mu}} - e^{ -
\frac{i}{2}\frac{\partial}{\partial \mu}})
\langle (H^{r})_{W} \rangle _{R=1}  ,
\label{form}
\ee
where $\langle (H^{r})_{W} P\rangle $ is given by (\ref{hwp}) and
$\langle (H^{r})_{W} \rangle _{R=1}$ by
\be
\langle (H^{r})_{W} \rangle _{R=1} = \frac{1}{r+1}
res\left( D^{r} (\mu D^{-1} -x)^{r+1} \right) _{W} ,
\label{tjohej}
\ee
where the right hand side is given by (\ref{hr}).
The necessary tools for this proof are the elementary
formulae
$e^{z\frac{\partial}{\partial x}} f(x) = f(x+z)$ and
\be
\left( x+\frac{1}{2}\right) _{n}
- \left( x-\frac{1}{2}\right) _{n} = n
\left( x+\frac{1}{2}\right) _{n-1}.
\label{der}
\ee
The right-hand-side of (\ref{form})
then becomes, using (\ref{hr}),
\be
-\frac{1}{r+1} \sum _{k=0}^{r} 2^{-k} \left( \begin{array}{c}
r+1 \\ k
\end{array} \right) \frac{r!}{(r-k)!} (r-k+1)i^{r-1}(-1)^{r-k}
\left( i\mu +\frac{1}{2}\right) _{r-k}.
\label{svar}
\ee
A generalization of (\ref{der})
\be
\sum _{m=0}^{n} \left( \begin{array}{c}
n \\ m
\end{array} \right) (x-m)_{r} (-1)^{m} = r(r-1)...(r-n+1) (x)_{r-n}
\ee
transforms (\ref{svar}) into
\be
\frac{i^{r+1}}{r+1} \sum _{k=0}^{r} \sum _{m=0}^{k} 2^{-k} \left(
\begin{array}{c}
r+1 \\ k
\end{array} \right) \left( \begin{array}{c}
k \\ m
\end{array} \right) (r-k+1)\left(
i\mu +\frac{1}{2}-m\right) _{r}
(-1)^{r-k+m}   .
\ee
Exchange the sums and use $\sum _{k=0}^{n} \left( \begin{array}{c}
n \\ k
\end{array} \right) (-2)^{-k} = 2^{-n}$. The result of this is
(\ref{hwp}) and
equality (\ref{tjohej}) is proven!
This confirms that the approach of the previous section reproduces
the results of the $c=1$ matrix model at $R=1$ for all genera.

\section{THE DEFORMED MATRIX MODEL}

\subsection{Introduction}

I have now come to the case of the deformed matrix model.
Let me quickly go through the basics of this model as it was
introduced in \cite{jev} and developed in subsequent papers
[2-6]. The deformed matrix model is obtained by adding a piece
$M/2x^{2}$ to the matrix model potential. The potential becomes
\be
-\frac{1}{2\alpha '} x^{2} +\frac{M}{2x^{2}}  ;
\ee
$M$  will be positive. The position of the Fermi level
is still measured in terms of its deviation from zero, i.e. $\mu$.
However, it is now
possible to define a double scaling limit, even when $\mu =0$. One then
needs to keep
$\hbar /M^{1/2}$ fixed,
which will be the string coupling constant.

Special cases of tachyon correlation functions have been calculated
in several papers, [2,3,5,6]. In \cite{dual} the general formula
(in the case with $N-1$ tachyons of the same chirality)
up
to genus one
was calculated to be, for zero $\mu$,
$$
\langle T_{p_{1}}...T_{p_{N-1}}T_{-p} \rangle
$$
$$
=
(N-3)!! p(p-2)...(p-(N-4))\prod _{i=1}^{N-1} p_{i}
\bigg[ M^{p/2 -N/2 +1}  -(p-(N-2))
$$
\be
\times \left(
\frac{p^{2}+(N-1)\sum _{i=1}^{N-1} p_{i}^{2}
-2(N-2)p -4N+1}{24} -\frac{(N-1)}{24R^{2}}\right)
M^{p/2-N/2} +... \bigg]     ,
\label{tmcorr}
\ee
when normalized to collective field theory.
At genus zero it is easy to use this formula to write down the
tachyon $N$-point function also for non-zero $\mu$. In this case,
not only $\hbar /M^{1/2}$ is kept constant in the double scaling limit,
but also $\hbar /\mu$. Both are proportional to
the string coupling constant.
One finds
\be
\langle T_{p_{1}}...T_{p_{N-1}}T_{-p} \rangle
= \frac{1}{p} \frac{\partial ^{N-2}}
{\partial \mu ^{N-2}} (M +\mu ^{2})^{p/2}    .
\label{tcorr}
\ee

\subsection{An Ansatz}

The purpose of this subsection is to reproduce the above results using
a generalization of the framework exemplified in previous
sections.
The formula I will generalize to non-zero $M$ is not (\ref{1t}) but rather
\be
\langle T_{1}T_{-p}\rangle =\frac{\partial}{\partial x} \langle
T_{-p} \rangle =\frac{-1}{p} res(W^{p})  .
\label{tt}
\ee
I will propose that the following expression for $W$ should be put
into (\ref{tt}):
\be
W = \left( M D^{-2} +\left(
\mu D^{-1} - x +\sum _{i=2}^{\infty} t_{i}
D^{i-1}\right) ^{2}\right) ^{1/2}    .
\label{supgen}
\ee
At $M=0$ we get (\ref{sup}) back. Let me now verify that
(\ref{supgen}) gives the correct correlation functions
on the sphere. In this case the ordering inside of (\ref{tt}) is trivial.
After integration of (\ref{tt}) with respect to $x$ it is obtained that
\be
\langle T_{-p} \rangle = \frac{1}{p} \sum _{l=0}^{p/2} \left(
\begin{array}{c}
p/2 \\ l
\end{array} \right)
\frac{1}{p-2l+1} M^{l} res \left( D^{-2l}
\left( \mu D^{-1} - x +\sum _{i=2}^{\infty} t_{i} D^{i-1}\right)
^{p-2l+1} \right)   .
\label{jotack}
\ee
After derivations with respect to the parameters $t_{k}$, equation
(\ref{tcorr}) is indeed obtained.

At this point some clarifications are needed. Throughout this
section I will assume the following rule of integration
\be
\int dx
\left( \mu D^{-1} - x +\sum _{i=2}^{\infty} t_{i} D^{i-1}\right)
^{k} =
\frac{-1}{k+1}
\left( \mu D^{-1} - x +\sum _{i=2}^{\infty} t_{i} D^{i-1}\right)
^{k+1} .
\ee
The basic formula that defines correlation functions of
topological gravity is often taken to be a two-point function
of the form
(\ref{tt}). The integration that leads to a one-point function
of the form (\ref{1t}) is justified using the string equation, see
e.g. \cite{wit3}. This is equivalent to the rule above.

Let me proceed to show that the agreement continues beyond the sphere.
In this case, however, one must be careful with the
ordering as determined by (\ref{supgen}). $W^{p}$
is given, for even $p$, as an
expansion in which all orderings of the factors $M D^{-2}$
and $(\mu D^{-1} -x+\sum _{i=2}^{\infty} t_{i} D^{i-1} )^2$
appear with equal weight. I will consider the two-point function to
genus 1. The only terms in (\ref{jotack}) that will contribute are the
ones with at most $x^{2}$ and just one power of $t_{m}$. Hence we
should pick $p-2l =2$, which gives
\be
\frac{1}{3p} \sum _{r=0}^{p/2-1}
  res \left( (M D^{-2})^{r}
\left( \mu D^{-1} - x +\sum _{i=2}^{\infty} t_{i} D^{i-1}\right)
^{3} (M D^{-2})^{p/2 -1-r}
\right)  . \label{tt1}
\ee
We must then pick out the $x^2$ and $t_{m}$ terms, i.e.
\be
(-x+t_{m} D^{m-1} )^{3} \rightarrow 3t_{m}x^{2} D^{m-1} +
3\hbar t_{m} x(m-1) D^{m-2} +\hbar ^{2} t_{m}
(m-1)(m-2)            ,
\ee
and then make the substitution in (\ref{tt1}). The $x$ and
$x^2$ are commuted
to the left, and put to zero. The term
proportional to $\hbar ^2$ is then
\be
\frac{1}{3p}
\sum _{r=0}^{p/2-1} M ^{p/2-1} res \left( (6r(2r+1)
 -6r(m-1)+(m-1)(m-2))D^{m-p-1} \right)  .
\ee
This implies $m=p$, i.e. momentum conservation, and the
final result is
\be
\frac{1}{p} M ^{p/2} -\frac{1}{12}M ^{p/2-1}(p-2)(p+2)
\ee
if the genus zero term also is included.
Let us now go back to the matrix model result (\ref{tmcorr}).
Specializing this formula to $N=2$ gives
\be
\frac{M ^{p/2}}{p}
 -\frac{1}{24}\left(
 2p^{2}-7-\frac{1}{R^{2}}\right) M ^{p/2-1} +...
\ee
So, at $R=1$ the answer agrees with the result of the calculation
based on (\ref{supgen})!

In the appendix, I give a similar calculation for the four-point function.

\subsection{A D-series Superpotential}

The $W$ of (\ref{supgen}) looks very awkward,
due to the square root. This
can, however, be dealt with. In the deformed matrix model it is
well known that only half of the discrete tachyon states appear,
the ones with even momenta. This means, for instance,
that the momentum $p$ in
(\ref{tt}) is always even. Therefore it is natural to relabel all the
states by replacing the primaries $\phi _{n}$ by $\phi _{n/2}$
and descendants $\sigma _{m}$  by $\sigma _{m/2}$.
As a consequence it is rather the square of (\ref{supgen}) that
should be thought of as the generalized superpotential, i.e.
\be
W = M D^{-2} +\left( \mu D^{-1} - x +\sum _{i=2}^{\infty} t_{i}
D^{i-1}\right) ^{2}   .
\label{nysup}
\ee
This still does look a bit strange. However, I will now show that
this expression can be derived from a D-series superpotential.
The D-series has been considered previously in the context of the
traditional topological field theories \cite{dvv,eguchi2}. There
the D-series potential
\be
X^{r} +\frac{1}{2}X Y^{2}
\ee
was considered. I now claim that the deformed matrix model
is based on the superpotential
\be
W = MX^{-1} -X Y^{2} +\sum _{i=0}^{\infty} 2t_{2i} YX^{i}   .
\ee
Following \cite{dvv,eguchi2} I integrate out the $Y$ field (this is
possible since it
appears only quadratically) using its equation of motion
\be
Y=\sum _{i=0}^{\infty} t_{2i} X^{i-1} .
\ee
There is also a Jacobian which is taken care of by the change of
variables $X=Z^{2}$, see \cite{dvv,eguchi2}.
The result of these manipulations is
\be
W = MZ^{-2} +\left( \sum_{i=0}^{\infty} t_{2i} Z^{2i-1}\right)
^{2}    ,
\label{dpot}
\ee
which is (\ref{nysup}) after the proper identifications!
Hence I conclude that {\it the deformed matrix model is
given by a D-series Landau-Ginzburg potential $MX^{-1} -XY^{2}$
with the
deformations $YX^{k}$}. The $YX^{k}$ are to be identified with the
special tachyons $T_{2k}$.

One should note that the coupling of the $T_{1}$ tachyon, $x$, is put
to zero in (\ref{dpot}). This is just as it should,
since only even-momentum special
tachyons exist in the deformed model. However, as is
clear from the previous subsection, it still plays a crucial role in
the higher genus calculations.

\section{CONCLUSIONS}

We have seen how the results of the matrix model can be obtained
for all genera
by using a topological field theory approach. I have
also shown how a change in the prescription is capable of
reproducing the results of the deformed matrix model.

It seems as if
there are some rather remarkable connections between the
matrix models and the Landau-Ginzburg formulation that is
fully revealed only in the light of the deformed matrix model.
The suggestive derivation of the deformed matrix model from a
D-series superpotential should be a clue to the understanding of
the physics behind the model. Perhaps this will teach us enough about
target space string physics for us to
finally understand the
elusive matrix model black hole.

\section*{Acknowledgements}

I would like to thank E. Verlinde and W. Lerche for discussions.

\section*{Appendix}

In this appendix I will calculate the four-point correlation
functions $\langle T_{m} T_{m} T_{m} T_{-p} \rangle$.
The relevant terms of (\ref{tt}) are
$$
\frac{1}{p} \int dx \sum _{s=0} ^{p/2-2} \sum _{r=0}^{p/2-2-s}
  res \left( (M D^{-2})^{r}
(\mu D^{-1} - x +t_{m} D^{m-1}) ^{2}
\right.
$$
\be
\left.
\times (M D^{-2})^{s}
(\mu D^{-1} - x +t_{m} D^{m-1}) ^{2}
(M D^{-2})^{p/2 -2-r-s} \right)      .
\ee
Collect the factors of $\mu D^{-1} - x +t_{m} D^{m-1}$ and perform
the $x$ integration to obtain
$$
\frac{1}{p}  \sum _{s=0} ^{p/2-2} M ^{p/2-2}
\sum _{r=0}^{p/2-2-s}
 res \left( \frac{1}{5} D^{-2r}
(\mu D^{-1} - x +t_{m} D^{m-1}) ^{5}
D^{-p+4+2r}
\right.
$$
$$
-4\hbar s \frac{1}{4} D^{-2r}
(\mu D^{-1} - x +t_{m} D^{m-1}) ^{4}D^{-p+3+2r}
$$
\be
\left.
+(4s^{2}+2s)\hbar ^{2} D^{-2r}
(\mu D^{-1} - x - t_{m} D^{m-1}) ^{3}
D^{-p+2+2r} \right)  .
\ee
The first term becomes
$$
\frac{1}{5p} M ^{p/2-2}t_{m}^{3} \sum _{s=0} ^{p/2-2}
\sum _{r=0}^{p/2-2-s}
res \left( D^{-2r}(10x^{2}D^{3m-3}+30\hbar x(m-1)D^{3m-4}
\right.
$$
$$
\left.
+ \hbar ^{2} (m-1)(25m-35) D^{3m-5})D^{-p+4+2r} \right)
$$
$$
=\frac{1}{p} M ^{p/2-2}t_{m}^{3} \sum _{s=0} ^{p/2-2}
\sum _{r=0}^{p/2-2-s} (4r(2r+1)-12r(m-1)+(m-1)(5m-7))
\delta _{3m,p}
 $$
\be
= M^{p/2-2}t_{p/3}^{3} (p-2)\left(
\frac{p^{2}}{36} -\frac{p}{12} -\frac{1}{8}\right)           .
\ee
The second term becomes
$$
-\frac{1}{p} M ^{p/2-2}t_{m}^{3} \sum _{s=0} ^{p/2-2}
\sum _{r=0}^{p/2-2-s}
 res \left( \hbar sD^{-2r}(4xD^{3m-3}+6\hbar (m-1)D^{3m-4})
D^{-p+3+2r} \right)
$$
$$
=
-\frac{1}{p} M ^{p/2-2}t_{m}^{3} \sum _{s=0} ^{p/2-2}
\sum _{r=0}^{p/2-2-s}
(6s(m-1)-8rs)  \delta _{3m,p}
$$
\be
= -M^{p/2-2} t_{p/3}^{3} \frac{p}{48}(p-2)(p-4)   .
\ee
Finally, the third term is calculated to be
\be
M^{p/2-2}t_{p/3}^{3}\frac{p}{144}(p-2)(p-4) .
\ee
Adding these contributions together, and taking the $t_{p/3}$
derivatives, gives the answer
\be
M^{p/2-2} \frac{p-2}{24}(2p^{2}-4p-18)    .
\ee
Quite remarkably, this agrees with (\ref{tmcorr}) at $R=1$.

\end{document}